\documentclass[prd,prerprint,showpacs,showkeys,preprintnumbers,amsmath,amssymb,floatfix]{revtex4}

\usepackage[utf8]{inputenc}
\usepackage{hyperref}

\usepackage{dcolumn}
\usepackage{bm}
\usepackage{graphicx}
\usepackage{subfigure}

\def\be{\begin{equation}}
\def\ee{\end{equation}}
\def\bestar{\begin{equation*}}
\def\eestar{\end{equation*}}

\begin{document}

\title{Thermodynamical properties of a noncommutative anti-de Sitter-Einstein-Born-Infeld spacetime from gauge theory of gravity}

\author{Rom\'an Linares}
\email{lirr@xanum.uam.mx}
\affiliation{Departamento de F\'{i}sica \\ Universidad Aut\'{o}noma Metropolitana - Iztapalapa\\
                  Av. San Rafael Atlixco 186, C.P. 09340, México City, M\'exico}

\author{Marco Maceda}
\email{mmac@xanum.uam.mx}
\affiliation{Departamento de F\'{i}sica \\ Universidad Aut\'{o}noma Metropolitana - Iztapalapa\\
                   Av. San Rafael Atlixco 186, C.P. 09340, México City, M\'exico}                   
                                  
\author{Oscar S\'anchez-Santos}
\email{osanchezs@up.edu.mx}
\affiliation{Universidad Panamericana \\
Augusto Rodin 498, C.P. 03920, México City, México}   
\affiliation{Instituto de Educación Media Superior de la Ciudad de México \\
Av. División del Norte No. 906, Col. Narvarte Poniente, C. P. 03020 Del. Benito Juárez, México City, México.}

\date{\today}

\begin{abstract}
We construct a deformed anti-de Sitter-Einstein-Born-Infeld black hole from noncommutative gauge theory of gravity and determine the metric coefficients up to second order on the noncommutative parameter. We analyse the modifications on the thermodynamical properties of the black hole due to the noncommutative contributions, and we show that noncommutativity has as a direct consequence, the removal of critical points.
\end{abstract}


\maketitle

\section{Introduction}
\label{intro}

In the past years, several approaches have been proposed to explore fundamental interactions in nature at energies beyond those attained at the laboratory. Gauge theories, for example, provide many helpful insights into the nature of particle interactions, and they have been generalised to various contexts. In particular, noncommutative gauge theories using the Seiberg-Witten map~\cite{Seiberg:1999vs} have broad applications into the analysis of the scattering process of fundamental particles at high energies~\cite{Moller:2004qq,Alboteanu:2007bp,Horvat:2011qn,KumarDas:2013twa,Selvaganapathy:2016jrl}. On the other hand, loop quantum gravity, string theory and noncommutative geometry are a few of several approaches that exist aiming to elucidate the structure of spacetime taking into account quantum effects; the link between the string theory and noncommutative geometry has attracted much attention since first noticed~\cite{Ferrari:2013pq,Stern:2014uea,Matsumoto:2014gwa,Blair:2014kla,Walters:2015bda}. The ultimate goal of all these formalisms is to provide a detailed picture of our universe at high energies.

At the Planck scale, noncommutativity manifests itself through discretisation of spacetime and quantum gravitational effects are noticeable~\cite{Brandenberger:2007zz}. A procedure to analysing these effects is to use commutation relations among spacetime coordinates. The implementation of a star product provides a simple realisation of these commutation relations and at the same time, generalises the standard pointwise multiplication of functions. The star product defined on a manifold encodes the quantum structure of spacetime and its discretisation naturally. 

One of the implications of a noncommutative structure of spacetime is that the standard point-like nature of particles becomes replaced by a fuzzy picture; objects are no longer localised ~\cite{Madore:1997ta,Madore:1999bi}. This fundamental difference with respect to the classical scenario leads to the prediction of new effects susceptible of experimental observation in particle physics, gravity and cosmology~\cite{Moffat:2000gr,Moffat:2000fv,Aschieri:2005yw,Aschieri:2005zs,MullerHoissen:2007xy,Pimentel:2004jv,Pimentel:2006at,Ohl:2009pv}. In all of these scenarios, the fuzzy structure of the sources affects the spacetime with non-trivial consequences. 

Furthermore, noncommutative gravity produces quantum corrections and deviations of the classical results that may be used to set bounds on the noncommutative parameter. In particular, the thermodynamical properties of noncommutative spacetimes may be useful in understanding the evolution of black holes when a minimal length scale exists~\cite{Dvali:2010gv,Spallucci:2014kua}. 

As the Seiberg-Witten map gives a straightforward construction of a noncommutative gauge theory from a commutative one employing a star product, it is natural to ask how this procedure works using the formulation of gravity as a gauge theory. Fortunately enough, gauge theories of gravity have a long history in physics and admit an elegant formulation in the language of fibre bundles~\cite{Cho:1975sw,Chang:1975fr,Mansouri:1976df,MacDowell:1977jt}. These models have as a basis a gauge theory of gravity with de Sitter group as their local symmetry group~\cite{Zet:2003bv} at the classical level. The classical spacetimes obtained in this way are then generalised to the noncommutative framework by a perturbative approach based on the Seiberg-Witten map; such theories of noncommutative gauge-gravity have been formulated and extensively studied in the last years~\cite{Chaichian:2007we,Chaichian:2007dr,Zet:2009zz}.

Even if models of pure gravity are exciting {\it per se}, either classical or noncommutative, the general situation where other interactions are present gives valuable information on several classes of phenomena. For example, in active galaxy nuclei, if an electromagnetic field exists, it should interact with itself in a nonlinear way, and the motion of charged test particles provides a way to quantify differences in comparison when the nonlinearity is absent. 

It is then necessary for a complete picture of fundamental interactions to include at some point, nonlinear electrodynamics. In this regard, Born-Infeld (BI) electrodynamics~\cite{Born:1934ji,Born:1934gh} and the corresponding gravitational solution, the standard Einstein-Born-Infeld (EBI) metric~\cite{GarciaD2007}, have a broad variety of applications and extensions, from inflation and branes to AdS and dilaton black holes.  For these reasons, in this paper, we use the gauge formulation of gravity to construct a noncommutative anti-de Sitter-Einstein-Born-Infeld (AdSEBI) black hole. A thermodynamical analysis of the solutions, based on the equation of state and Gibbs function, may reveal the existence or not of new critical points and new phase transitions induced by noncommutativity. 

The paper is organised as follows: In Sec.~\ref{secc:2}, we review the formalism of gauge theory of gravity based on the Sitter group, and use this approach there to solve explicitly the field equations for the specific case of AdSEBI spacetime; we recover the well-known EBIon solution with a cosmological constant when the BI parameter vanishes. We discuss then the deformation technique of gauge-gravity theory based on the Seiberg-Witten map and apply it to the commutative AdSEBI black hole in Sec.~\ref{secc:4} to define the noncommutative counterpart. The analysis of the noncommutative effects on the thermodynamical properties of the AdSEBI black hole is discussed in Sec.~\ref{secc:6}; there we show that critical points may be removed. Finally, in the Conclusions section, we give some remarks and perspectives about future work. 

\section{Gauge theory of gravity}
\label{secc:2}

We review briefly the main ingredients of gauge theory of gravity with de Sitter group as a local symmetry~\cite{Zet:2003bv,Blagojevic:2002du,Blagojevic:2013xpa}. The starting point is a 4-dimensional Minkowski metric in spherical coordinates
\begin{equation}\label{EC1}
ds^2=dt^2-dr^2-r^2\bigl(d\theta^2+\sin^2\theta d\phi^2\bigl).
\end{equation}  
This spacetime is the base manifold where the gravitational  field will be described by gauge field potentials $h^A_\mu, \mu = 0, \dots, 3, A = 1, \dots, 10$. The gauge potentials depend on the coordinates of the base manifold and are split into two sets: four tetrads $e^a$ with components $e^a_{\mu}(x)$, and six spin connections with components $\omega^{ab}_{\mu}(x)$; the latter possess the standard skew property $\omega^{ab}_{\mu}(x) = -\omega^{ba}_{\mu}(x)$. 

Using the tetrad and the spin connections, we define the following antisymmetric strength tensors 
\begin{equation}\label{EC2}
F^a_{\mu\nu} := \partial_{\mu}e^a_{\nu}-\partial_{\nu}e^a_{\mu}+\bigl(\omega^{ab}_{\mu}e^c_{\nu}-\omega^{ab}_{\nu}e^c_{\mu}\bigl)\eta_{bc},
\end{equation}  
\begin{eqnarray}\label{EC3}
F^{ab}_{\mu\nu} &:=& \partial_{\mu}\omega^{ab}_{\nu}-\partial_{\nu}\omega^{ab}_{\mu}+\bigl(\omega^{ac}_{\mu}\omega^{db}_{\nu}-\omega^{ac}_{\nu}\omega^{db}_{\mu}\bigl)\eta_{bc}
\nonumber \\[4pt]
&&+4\lambda^2\bigl(e^b_\mu e^a_\nu - e^a_\mu e^b_\nu \bigl).
\end{eqnarray}
The action associated with these gauge gravitational fields is 
\begin{equation}\label{EC4}
S_{g}={1\over 16\pi G}\int d^4 x \, e \,F,
\end{equation}
where $e := \mbox{det}(e^a_{\mu})$  and 
\begin{equation}\label{EC5}
F := F^{ab}_{\mu\nu}\bar e^{\mu}_a\bar e^{\nu}_b.
\end{equation}
Here the elements $\bar e^{\mu}_a(x)$ are the inverse elements of the tetrad components $e^a_{\mu}(x)$, i.e. $\bar e^{\mu}_a \bar e^a_\nu = \delta^\mu_\nu$, etc. 

As we are dealing with gauge gravitational fields, an electromagnetic gravitational potential $A_{\mu}(x)$ is also created; its corresponding action is given by 
\begin{equation}\label{EC6}
S_{em}=-{1\over 4Kg^2}\int d^4 x \, e \, A^a_{\mu}\bar A^{\mu}_a,
\end{equation}     
where the quantities $A^a_{\mu}$ and $\bar A^{\mu}_a$ are defined as 
\begin{equation}\label{EC7}
A^a_{\mu} := A^{\nu}_{\mu}e^a_{\nu}, \quad A^{\nu}_{\mu} := \bar e^{\nu}_a\bar e^{\rho}_b\eta^{ab}A_{\mu\rho}, \quad 
\bar A^{\mu}_a := A^{\mu}_{\nu}\bar e^{\nu}_a,
\end{equation}     
in terms of the electromagnetic field tensor 
\begin{equation}\label{EC8}
A_{\mu\nu} := \partial_{\mu}A_{\nu}-\partial_{\nu}A_{\mu}.
\end{equation}
In the above expressions, $g$ represents a coupling constant and $K$ is a normalisation constant that will be chosen later to simplify certain expressions.
 
The full action of the model takes into account the gauge fields and the electromagnetic field 
\begin{equation}\label{EC9}
S=\int d^4x \, e\biggl[{1\over 16\pi G}F-{1\over 4Kg^2}A^a_{\mu}\bar A^{\mu}_a\biggl].
\end{equation}     
Variation of this action with respect to the tetrad components $e^a_{\mu}(x)$ gives the following field equations  
\begin{equation}\label{EC10}
F^{a}_{\mu}-{1\over 2}Fe^a_{\mu}=8\pi GT^{a}_{\mu},
\end{equation}     
where $F^{a}_{\mu} := F^{ab}_{\mu\nu}\bar e^{\nu}_b$ and the energy-momentum tensor is given by
\begin{equation}\label{EC11}
T^{a}_{\mu}={1\over Kg^2}\biggl(A^b_{\mu}A^a_{\nu}\bar e^{\nu}_b - \frac 14 A^b_{\nu}A^{\nu}_b e^a_{\mu}\biggl).
\end{equation}     
When considering additional fields interacting with gravity, the corresponding energy-momentum tensors should be included on the right hand side of Eq.~(\ref{EC10}). 

On the other hand, the field equations for the gravitational gauge potentials $\omega^{ab}_{\mu}(x)$ are 
\begin{equation}\label{EC12}
F^a_{\mu\nu}=0.
\end{equation}     
Physically, this condition amounts to the absence of torsion in the theory as it is the situation in General Relativity. 

\subsection{Commutative AdSEBI spacetime from gauge theory of gravity} 
\label{secc:3}

To illustrate the previous formalism of gauge-gravity~\cite{Zet:2003bv,Zet:2004xn,Zet:2007zz,Babeti:2014vya} for classical spacetimes, we use it now to obtain a commutative anti-de Sitter-Einstein-Born-Infeld black hole solution. The starting point is a gravitational gauge field with spherical symmetry given by the following Ansatz
\begin{eqnarray}\label{EC13}
&&e^{0}_{\mu}=(A, 0, 0, 0),\hspace{0.5cm}e^{1}_{\mu}=\biggl(0, {1\over A}, 0, 0\biggl),\hspace{0.5cm}
\nonumber \\[4pt]
&&e^{2}_{\mu}=(0, 0, r, 0),\hspace{0.5cm}e^{3}_{\mu}=(0, 0, 0, r\sin\theta),
\end{eqnarray}     
together with the spin connections
\begin{eqnarray}\label{EC14}
&&\omega^{01}_{\mu}=(U, 0, 0, 0),\hspace{0.5cm}\omega^{02}_{\mu}=\omega^{03}_{\mu}=0,\hspace{0.5cm}\omega^{12}_{\mu}=(0, 0, A, 0),
\nonumber \\[4pt]
&&\omega^{13}_{\mu}=(0, 0, 0, A\sin\theta),\hspace{0.5cm}\omega^{12}_{\mu}=(0, 0, 0, \cos\theta),
\end{eqnarray}     
where $A$ and $U$ are functions of the radial coordinate $r$. On the other hand, the density Lagrangian of the BI electrodynamics is chosen as 
\be
L_{BI} = \frac 2{b^2} \left( 1 - \sqrt{1 + b^2 F_{\mu\nu} F^{\mu\nu}} \right),
\ee
where $b$ is the BI parameter; in the limit when $b\to 0$, we recover standard Maxwell's electrodynamics. From the above Lagrangian, we obtain straightforwardly the components of the energy-momentum tensor of the BI electrodynamics. In consequence, the field equations for the tetrad elements $e^a_{\mu}$ as deduced from Eqs.~(\ref{EC10}) including the BI electromagnetic contributions are 
\begin{widetext}
\begin{eqnarray}
\label{EC15}
&&\biggl({-2rAA^\prime+1-A^2+12\lambda^2r^2\over r^2}\biggl)A=\biggl({\sqrt{r^4+b^2Q^2}\over r^2b^2}-{1\over b^2}\biggl)A,
\\[4pt]
\label{EC16}
&&\biggl({2rU+1-A^2+12\lambda^2r^2\over r^2}\biggl){1\over A}=\biggl({\sqrt{r^4+b^2Q^2}\over r^2b^2}-{1\over b^2}\biggl){1\over A},
\\[4pt]
\label{EC17}
&&rU^\prime+U-AA^\prime+12\lambda^2r={r\over b^2}\biggl({r^2\over \sqrt{r^4+b^2Q^2}}-1\biggl),
\\[4pt]
\label{EC18}
&&\biggl({r^2U^\prime+rU-rAA^\prime+12\lambda^2r^2\over r^2}\biggl)r\sin\theta={r\sin\theta\over b^2}\biggl({r^2\over \sqrt{r^4+b^2Q^2}}-1\biggl).
\end{eqnarray}   
\end{widetext}  
We see that Eqs.~(\ref{EC17}) and~(\ref{EC18}) are the same field equation. We also notice that Eqs.~(\ref{EC15}) and~(\ref{EC16}) become the same field equation if $U = -AA^\prime = - (A^2/2)^\prime$; this last relation is also a consequence from Eq.~(\ref{EC12}). Therefore, the relevant field equations to solve are
\begin{eqnarray}
\label{EC19}
&&-{2AA^\prime\over r}+{1-A^2\over r^2}+12\lambda^2=-{1\over b^2}+{\sqrt{r^4+b^2Q^2}\over r^2b^2},
\\[4pt]
\label{EC20}
&&-\frac 12(A^2)^{\prime\prime}-{2AA^\prime\over r}+12\lambda^2=-{1\over b^2} + {r^2\over {b^2\sqrt{r^4+b^2Q^2}}}.
\end{eqnarray}   
If we now take the difference between Eqs.~(\ref{EC19}) and~(\ref{EC20}), we obtain the single equation 
\begin{equation}\label{EC21}
\frac 12 (A^2)^{\prime\prime}+{1-A^2\over r^2}={Q^2\over r^2\sqrt{r^4+b^2Q^2}},
\end{equation}   
or equivalently
\begin{equation}\label{EC23}
r^2 (A^2)^{\prime\prime}-2 A^2 + 2 = {2Q^2\over \sqrt{r^4+b^2Q^2}}.
\end{equation}  
The solution to Eq.~(\ref{EC23}) is 
\begin{eqnarray}\label{EC24}
A^2 &=& 1-{2M\over r}-{\Lambda\over 3}r^2+{2\over 3}{r^2\over b^2}\biggl(1-\sqrt{1+{r_0^4 \over r^4}}\biggl)
\nonumber \\[4pt]
&&+\frac 43 \frac {r_0^4}{b^2 r^2} {}_2 F_1 \left( \frac 14, \frac 12; \frac 54; -\frac {r_0^4}{r^4} \right),
\end{eqnarray}
where we have set $r_0^4 := b^2 Q^2$ and $\Lambda := -12\lambda^2$. In the limit $b \to 0$, we recover the black hole solution of the model studied in~\cite{Chaichian:2007dr}. 

\section{Noncommutative adSEBI black hole from gauge theory of gravity}
\label{secc:4}

Having as starting point a gauge theory of gravity, a natural generalisation of this theory to the noncommutative framework makes use of the Seiberg-Witten map~\cite{Seiberg:1999vs}. A deformation based on the Moyal-Groenewold star product~\cite{Groenewold:1946kp,Moyal:1949sk} makes use of constant noncommutative parameters $\Theta^{\mu\nu}$; they define the commutation relations among the spacetime coordinates providing the noncommutative structure of spacetime as
\begin{equation}\label{EC25}
[x^{\mu}, x^{\nu}]=i\Theta^{\mu\nu},
\end{equation}  
where  $\Theta^{\mu\nu}$ is an antisymmetric matrix. It follows then that the product of any two fields is given by the Moyal star product
\begin{equation}\label{EC26}
(f\star g)(x)=f(x)e^{{1\over 2}\Theta^{\mu\nu} \overleftarrow{\partial_{\mu}} \overrightarrow{\partial_{\nu}}} g(x). 
\end{equation} 

Along the lines of~\cite{Chaichian:2007we}, noncommutative corrections to commutative quantities are obtained as follows. First, 
the noncommutative gauge gravity fields $\hat\omega^{AB}_{\mu}(x, \Theta)$ are subject to the reality conditions 
\begin{eqnarray}\label{EC27}
&&\hat{\omega^{AB}}^\dagger_{\mu}(x, \Theta)=-\hat\omega^{BA}_{\mu}(x, \Theta), 
\nonumber \\[4pt]
&&\hat{\omega^{AB}}^\dagger_{\mu}(x, \Theta)^r=\hat\omega^{AB}_{\mu}(x, -\Theta)=-\hat\omega^{BA}_{\mu}(x, \Theta), 
\end{eqnarray}
where $\dagger$ denotes complex conjugation; these conditions guarantee that the noncommutative gauge fields are real. Then, we expand the fields $\hat{\omega^{AB}}^{\dagger}_{\mu}(x, \Theta)$ in powers of the noncommutative parameter $\Theta$,
\begin{eqnarray}\label{EC28}
\hat{\omega^{AB}}^{+}_{\mu}(x, \Theta) &=& \omega^{AB}_{\mu}(x, \Theta)-i\Theta^{\nu\rho}\omega^{AB}_{\mu\nu\rho}(x)
\nonumber \\[4pt]
&&+\Theta^{\nu\rho}\Theta^{\lambda\tau}\omega^{AB}_{\mu\nu\rho\lambda\tau}(x)+ \dots.
\end{eqnarray}  
The reality conditions, Eqs.(\ref{EC27}), after expansion become
\begin{eqnarray}\label{EC29}
&&\omega^{AB}_{\mu}(x)=-\omega^{BA}_{\mu}(x),\qquad \omega^{AB}_{\mu\nu\rho}(x)=\omega^{BA}_{\mu\nu\rho}(x),
\nonumber \\[4pt]
&&\omega^{AB}_{\mu\nu\rho\lambda\tau}(x)=-\omega^{BA}_{\mu\nu\rho\lambda\tau}(x).
\end{eqnarray}  
Using now the Seiberg-Witten map, we obtain straightforwardly the corrections up to second order; they are~\cite{Chaichian:2007we}  
\begin{eqnarray}\label{EC30}
\omega^{AB}_{\mu\nu\rho}(x) &=& {1\over 4}\{\omega_{\nu}, \partial_{\rho}\omega_{\mu}F_{\mu\nu}\}^{AB},
\nonumber \\[4pt] 
\omega^{AB}_{\mu\nu\rho\lambda\tau}(x) &=& {1\over 32}(-\{\omega_{\lambda},\partial_{\tau}\{\omega_{\nu}, \partial_{\rho}\omega_{\mu}F_{\mu\nu}\}\}
\nonumber \\[4pt]
&&+2\{\omega_{\lambda},\{F_{\tau\nu}, F_{\mu\rho}\} \} 
\nonumber \\[4pt]
&&- \{\omega_{\lambda},\{\omega_{\nu}, D_{\rho}F_{\mu\nu}+\partial_{\rho}F_{\tau\mu}\}\}
\nonumber \\[4pt]
&&-\{\{\omega_{\nu},\partial_{\rho}\omega_{\mu}+F_{\mu\lambda}\}, (\partial_{\tau}\omega_{\mu}+F_{\tau\mu})\}
\nonumber \\[4pt]
&&+2[\partial_{\nu}\omega_{\lambda},\partial_{\rho}(\partial_{\tau}\omega_{\mu}+F_{\tau\mu})])^{AB},
\end{eqnarray}  
where we use the following definitions
\begin{eqnarray}\label{EC31}
\{\alpha, \beta\}^{AB} &:=& \alpha^{AC}\beta^B_C+\beta^{AC}\alpha^B_C,
\nonumber \\[4pt]
[\alpha, \beta]^{AB} &:=& \alpha^{AC}\beta^B_C-\beta^{AC}\alpha^B_C,
\end{eqnarray}  
and 
\begin{equation}\label{EC32}
D_{\mu}F^{AB}_{\rho\sigma} := \partial_{\mu}F^{AB}_{\rho\sigma}+(\omega^{AC}_{\mu}F^{BD}_{\rho\sigma}+\omega^{BC}_{\mu}F^{DA}_{\rho\sigma})\eta_{CD}.
\end{equation}  
Notice that we only need to know the classical quantities to determine the noncommutative contributions to them. 
 
On the other hand, the noncommutative corrections up to second order in the noncommutativity parameter to the classical tetrad fields are~\cite{Chamseddine:2000si,Ulker:2007fm,Ulker:2012yk} 
\begin{equation}\label{EC33}
\hat e^a_{\mu}(x, \Theta)=e^a_{\mu}(x)-i\Theta^{\nu\rho}e^a_{\mu\nu\rho}(x)+\Theta^{\nu\rho}\Theta^{\lambda\tau}e^a_{\mu\nu\rho\lambda\tau}(x),
\end{equation}  
where 
\begin{equation}\label{EC34}
e^a_{\mu\nu\rho}(x)={1\over 4}\bigl[\omega^{ac}_{\mu}\partial_{\rho}e^{d}_{\mu}+(\partial_{\rho}\omega^{ac}_{\mu}+F^{ac}_{\rho\mu})e^{d}_{\nu}\bigl]\eta_{cd},
\end{equation} 
and
\begin{widetext}
\begin{eqnarray}\label{EC35}
e^a_{\mu\nu\rho\lambda\tau}(x) &=& {1\over 32}\biggl[2\{F_{\tau\nu},F_{\mu\rho}\}^{ab}e^{c}_{\lambda}-\omega^{ab}_{\lambda}\bigl(D_{\rho}F^{cd}_{\tau\mu}+\partial_{\rho}F^{cd}_{\tau\mu}\bigl)e^{m}_{\nu}\eta_{dm} - \{\omega_{\nu},(D_{\rho}F_{\tau\mu}+\partial_{\rho}F_{\tau\mu})\}^{ab}e^{c}_{\lambda} +\partial_{\nu}\omega^{ab}_{\lambda}\partial_{\rho}\partial_{\tau}e^{c}_{\mu}
\nonumber \\[4pt]
&&-\partial_{\tau}\{\omega_{\nu},(\partial_{\rho}\omega_{\mu}+F_{\rho\mu})\}^{ab}e^{c}_{\lambda} - \omega^{ab}_{\lambda}\partial_{\tau}\bigl(\omega^{cd}_{\nu}\partial_{\rho}e^{m}_{\mu}+(\partial_{\rho}\omega^{cd}_{\mu}+F^{cv}_{\rho\mu})e^{m}_{\mu}\bigl)\eta_{dm} -  2\partial_{\rho}\bigl(\partial_{\tau}\omega^{ab}_{\nu}+F^{ab}_{\tau\mu}\bigl)\partial_{\nu}e^{c}_{\lambda}
\nonumber \\[4pt]
&&-\{\omega_{\nu},\bigl(\partial_{\rho}\omega_{\lambda}+F_{\rho\lambda}\bigl)\}^{ab}\partial_{\tau}e^{c}_{\mu} - \bigl(\partial_{\tau}\omega^{ab}_{\mu}+F^{ab}_{\tau\mu}\bigl)\biggl(\omega^{cd}_{\nu}\partial_{\rho}e^{m}_{\mu}+\bigl(\partial_{\tau}\omega^{ab}_{\mu}+F^{ab}_{\tau\mu}\bigl)e^{m}_{\mu}\eta_{dm}\biggl)\biggl]\eta_{bc}.
\end{eqnarray}
\end{widetext}
The noncommutative metric can then be found from the expression~\cite{Chaichian:2007we}
\begin{equation}\label{EC36}
\hat g_{\mu\nu}={1\over 2}\eta_{ab}\bigl(\hat e^{a}_{\mu}\star {{\hat e}^{b\dag}}_{\nu}+\hat e^{b}_{\mu}\star {{\hat e}^{a\dag}}_{\nu}\bigl),
\end{equation} 
where $\dag$ denotes complex conjugation.


We now proceed to calculate the noncommutative metric for the EBI spacetime using the formalism just described; using Eqs.~(\ref{EC34}) and~(\ref{EC35}) together with Eqs.~(\ref{EC13}) and~(\ref{EC14}), we obtain the following expressions 
\begin{widetext}
\begin{eqnarray}\label{EC37}
\hat g_{11} &=& {1\over A^2}+{1\over 4}{A^{''}\over A}\Theta^2+{\cal O}(\Theta^4),
\nonumber \\[4pt]
\label{EC38}
\hat g_{22} &=& r^2+{1\over 16}(A^2+11rAA^{'}+16r^2{A^{'}}^2+12r^2AA^{''})\Theta^2+{\cal O}(\Theta^4),
\nonumber \\[4pt]
\label{EC39}
\hat g_{33} &=& r^2\sin^2\theta+{1\over 16}\biggl[4\biggl(2rAA^{'}-r{A^{'}\over A}+r^2AA^{''}+r^2{A^{'}}^2\biggl)\sin^2\theta+\cos^2\theta\biggl]\Theta^2+{\cal O}(\Theta^4),
\nonumber \\[4pt]
\label{EC40}
\hat g_{00} &=& -A^2-{1\over 4}(2rA{A^{'}}^3+rA^3{A^{'''}}+A^3A^{''}+r^2{A^{'}}^2+2A^2{A^{'}}^2+5rA^2A^{'}A^{''})\Theta^2+{\cal O}(\Theta^4),
\end{eqnarray} 
\end{widetext}
up to second order on the noncommutative parameter. In the above expressions, the function $A$ is given by Eq.~(\ref{EC24}); we also set $\Theta^{12} = -\Theta^{21} =: \Theta$ as the only non-vanishing values for the noncommutative parameters.  We recover the classical result for the components of the metric in the commutative limit $\Theta\rightarrow 0$. 

Using now the explicit expression for $A$, a straightforward calculation shows then that the coefficient $\hat g_{00}$ of the noncommutative metric tensor is
\begin{widetext}
\begin{eqnarray}\label{EC41}
\hat g_{00} &=& 1-{2M\over r}-{\Lambda r^2\over 3}+{2\over 3}{r^2\over b^2}\biggl(1-\sqrt{1+{b^2Q^2\over r^4}}\biggl)+{4Q^2f(r)\over 3r}+{1\over 16} \left\{ 2r \left[ {4\over 3b^2}-{4M\over r^3} -{4r^4-8b^2Q^2\over 3b^2r^2\sqrt{r^4+b^2Q^2}} \right.\right.
\nonumber \\[4pt]
&&\left. -{2\Lambda\over 3}+{8Q^2f(r)\over 3r^3} \right] \left[ {2M\over r^2}+{4r\over 3b^2}\biggl(1-\sqrt{1+{b^2Q^2\over r^4}}\biggl)-{2\Lambda r\over 3}-{4Q^2f(r)\over 3r^2} \right]
\nonumber \\[4pt]
&&+\left[ {2M\over r^2}+{4r\over 3b^2}\biggl(1-\sqrt{1+{b^2Q^2\over r^4}}\biggl)-{\Lambda r^2\over 3}-{4Q^2f(r)\over 3r^2} \right]^2 + 2\left[ 1-{2M\over r}+{2\over 3}{r^2\over b^2} \biggl( 1-\sqrt{1+{b^2Q^2\over r^4}}\biggl) \right.
\nonumber \\[4pt]
&&\left. \left.-{\Lambda r^2\over 3}+{4Q^2f(r)\over 3r} \right] \left[ {4\over 3b^2}+{8M\over r^3}-{4(4b^4Q^4+11b^2Q^2r^4+r^8)\over 3b^2r^2(r^4+b^2Q^2)^{3/2}}-{2\Lambda\over 3}-{16Q^2f(r)\over 3r^3} \right] \right\} \Theta^2+{\cal O}(\Theta^4).
\end{eqnarray}
\end{widetext}
where 
\begin{eqnarray}
f(r) &:=& {1\over 2\sqrt{bQ}}F\biggl[\arccos\biggl\{{r^2+bQ\over r^2+bQ}\biggl\}, {1\over\sqrt{2}}\biggl]
\nonumber \\[4pt]
&=& \frac 1{2 r} {}_2 F_1 \left( \frac 14, \frac 12, \frac 54, - \frac {b^2 Q^2}{r^4} \right).
\end{eqnarray}

\begin{figure}[htbp!]
\centering
\includegraphics[width=210pt]{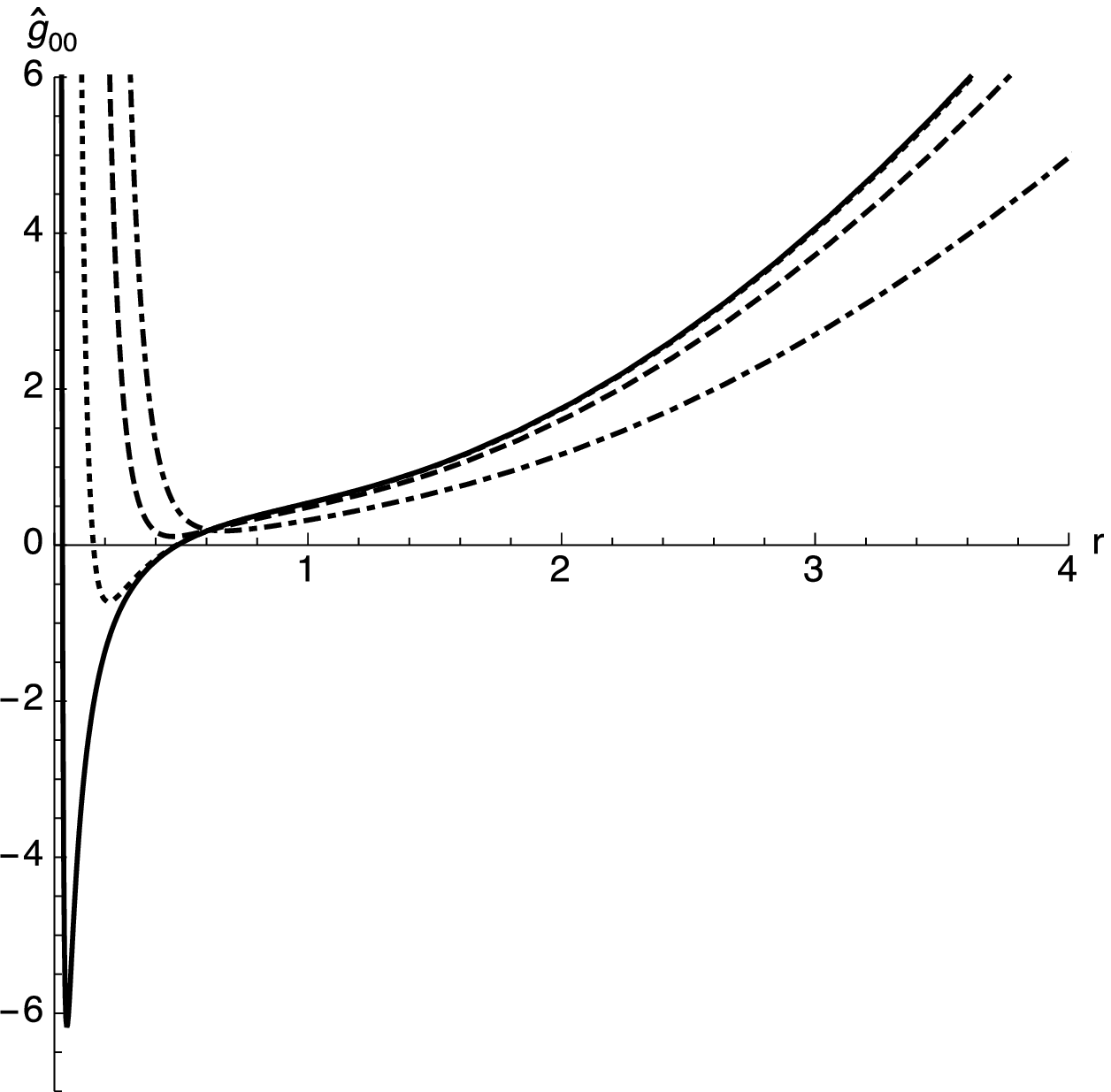}
\hfill
\includegraphics[width=210pt]{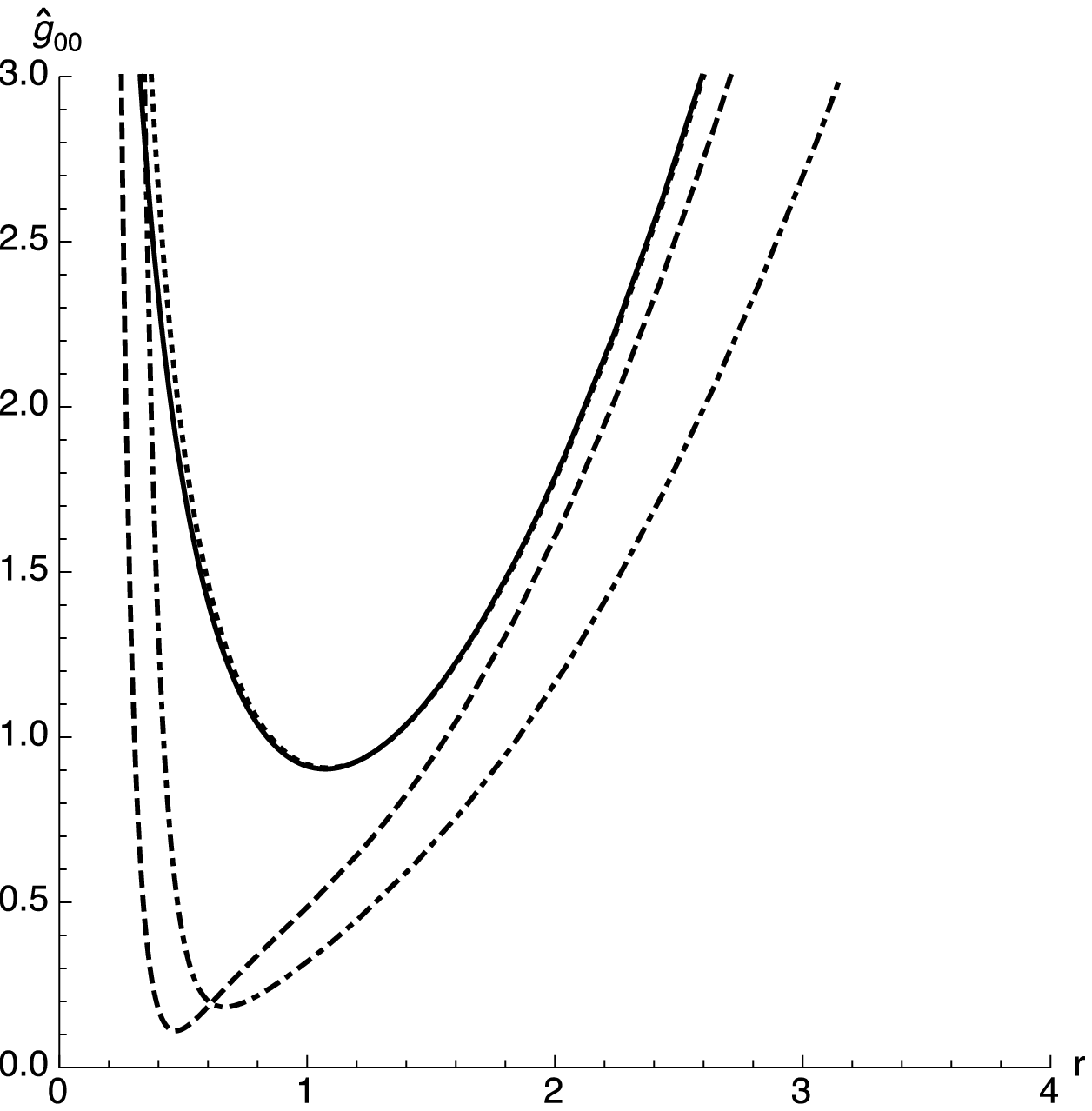}
\caption{Plots of $\hat g_{00}$ as a function of the radial variable $r$ for $\Theta = 0.01, 0.1, 0.4, 0.8$ (solid, dotted, dashed, dot-dashed line) with $Q=1, \Lambda = 3/2, M = 1$ and $b=1$ (left plot) and $b = 10$ (right plot). As in the classical case, the noncommutative AdSEBI black hole can develop two horizons for small values of the BI parameter $b$ that eventually collapse to only one; as $b$ becomes larger, the existence of horizons becomes harder.}
\label{Fig1}
\end{figure}

The explicit form of the remaining metric coefficients can be determined in a similar way. Once they are known, several quantities may be evaluated. In particular,  using the above expression for $\hat g_{00}$, we can determine the presence or absence of horizons. In Fig.~\ref{Fig1}, we plot $\hat g_{00}$ as a function of $r$; we see that the maximum number of horizons for the noncommutative AdSEBI black hole is two as in the commutative case. We also notice that as noncommutative effects are larger, an extremal black hole may arise with a unique horizon.

\section{Thermodynamics}
\label{secc:6}

The fact that a thermodynamical approach can be used to study black hole solutions is remarkable since it leads to a notion of temperature based on the surface gravity of the black hole. Quantum effects can also be incorporated in this picture, providing a more in-depth insight into certain aspects expected to exist in a quantum theory of gravity. 

In the following, we use the results of the previous sections to study the thermodynamics of the noncommutative AdSEBI black hole from gauge-gravity; the resulting expressions are direct generalisations of classical results. The first step in this direction is the calculation of the ADM mass. Using standard perturbation theory, the ADM mass up to second order in the noncommutative parameter is
\begin{widetext}
\begin{equation}\label{EC41b}
M_{ADM}={r_H\over 2}-{\Lambda r_H^3\over 6}+{r_H^3\over 3b^2}\biggl(1-\sqrt{1+{b^2Q^2\over r_H^4}}\biggl)+{4Q^2f(r_H)\over 3r_H}-{r_H^2\over 64}\biggl[{4\over r_H}\biggl({2Q^2\over\sqrt{r_H^4+b^2Q^2}}-1\biggl)
\end{equation}
\[
\biggl({1\over r_H}+{2r_H\over b^2}\biggl(1-\sqrt{1+{b^2Q^2\over r_H^4}}\biggl)-\Lambda r_H\biggl)+\biggl({1\over r_H}+{2r_H\over b^2}\biggl(1-\sqrt{1+{b^2Q^2\over r_H^4}}\biggl)-\Lambda r_H\biggl)^2\biggl]\Theta^2+{\cal O}(\Theta^4),
\]
\end{widetext}
where $r_H$ is the radius of the outer horizon. We see that the well-known classical result gives the leading order in this expression. On the other hand, the Hawking temperature becomes
\begin{eqnarray}\label{EC42}
T_{H} &=& {1\over 4\pi}{d\hat g_{00}\over dr}\biggl|_{r_H}
\nonumber \\[4pt]
&=&{1\over 4\pi r_H}\biggl[1+{2r_H^2\over b^2}-\Lambda r_H^2+{2r_H^2\over b^2}\biggl(1-\sqrt{1+{b^2Q^2\over r_H^4}}\biggl)\biggl]
\nonumber \\[4pt]
&&+ F_2(r) \Theta^2 + {\cal O}(\Theta^4),
\end{eqnarray}
where the explicit expression for the function $F_2(r)$, rather cumbersome, is provided in the Appendix. We expect that increasing orders in perturbation theory will give us more involved expressions for the noncommutative corrections.


We can also find the corrections to the pressure up to second order in the noncommutative parameter. We have the result
\begin{eqnarray}\label{EC42b}
P &=& {T_H\over v}-{1\over 2\pi v^2}+{b^2\over 4\pi}\biggl(1-\sqrt{1+{16b^2Q^2\over v^4}}\biggl) 
\nonumber \\[4pt]
&&+ F_3(r) \Theta^2 + {\cal O}(\Theta^4),
\end{eqnarray}
where the explicit expression for the function $F_3 (r)$ can be found in the Appendix. In Fig.~\ref{FigNivelJC.1}, we show several isotherms associated with the above equation of state; they have the particular form of isothermals of the van der Waals equation. For a fixed value of the BI parameter $b$, it is clear that for a small value of the noncommutative parameter $\theta$, a critical point exists, meanwhile for a larger value of $\theta$, the critical point is absent. Noncommutative effects make harder the presence of critical points. 

\begin{figure}[htp!]
\centering
\includegraphics[width=220pt]{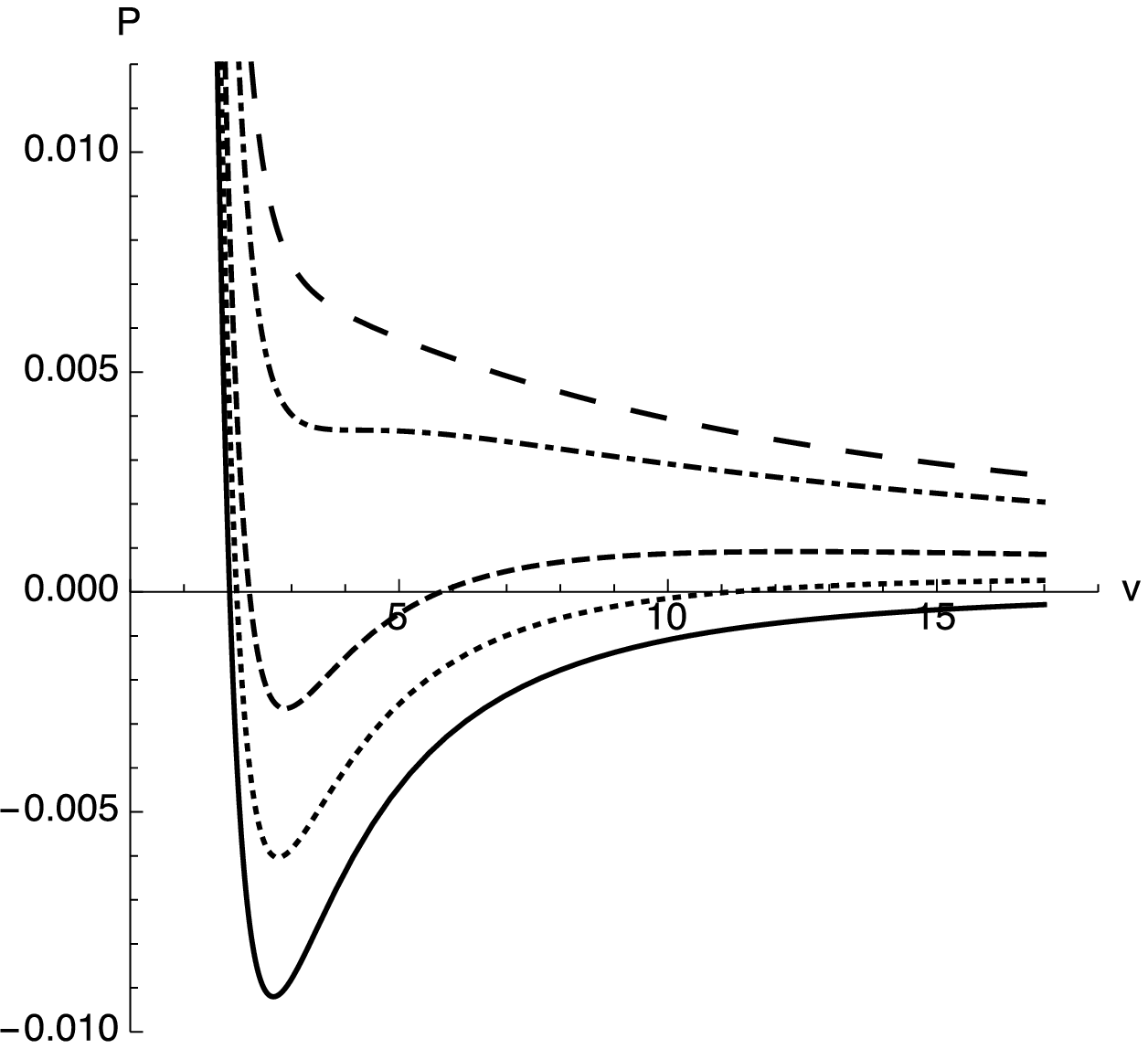}
\hfill
\includegraphics[width=220pt]{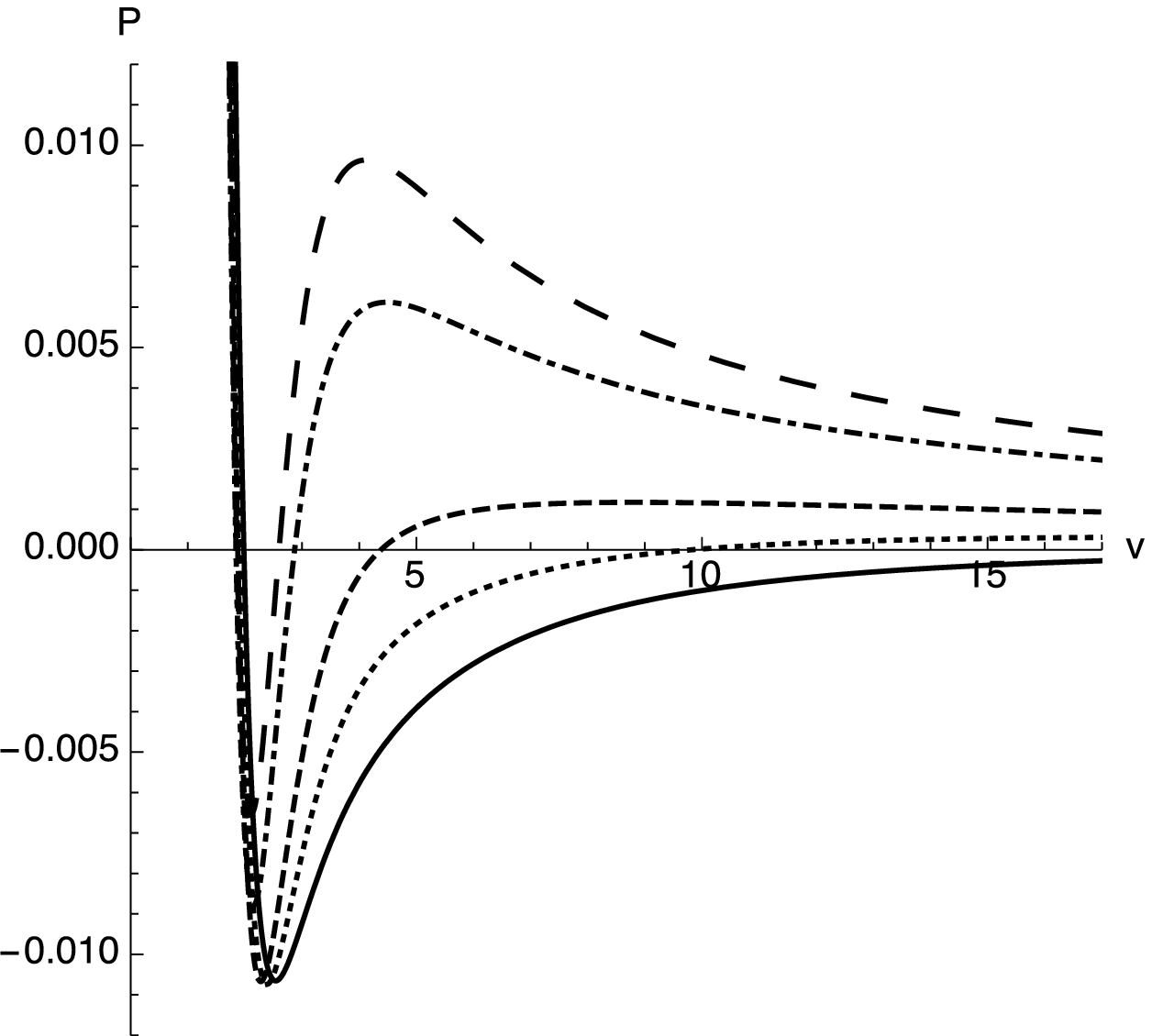}
\caption{$P-v$ diagram of noncommutative AdSEBI black hole. The corresponding values for the temperature are $T=0.0043, 0.04362, 0.02362, 0.01362$ and $0.05362$ (solid, dotted, dashed, dot-dashed and long dashing line) with $b=1, Q=1$ in both plots. For the upper plot, $\Theta=0.2$ and for the lower plot $\Theta=0.6$; noncommutativity makes harder the presence of critical points.}
\label{FigNivelJC.1}
\end{figure}


Besides the analysis of the equation of state of the noncommutative AdSEBI black hole, a discussion of critical points is achieved by calculating the corresponding Gibbs function $G$. In the commutative cases, both Schwarzschild and RN spacetimes have points in the $G-T_H$ plane where the derivative of $G$ as a function of the temperature $T_H$ is ill-defined, leading to the existence of phase transitions. 

\begin{widetext}

\begin{figure}[htbp]
\centering
\subfigure[] 
{
    \label{orbits:a}
    \includegraphics[width=0.3\textwidth]{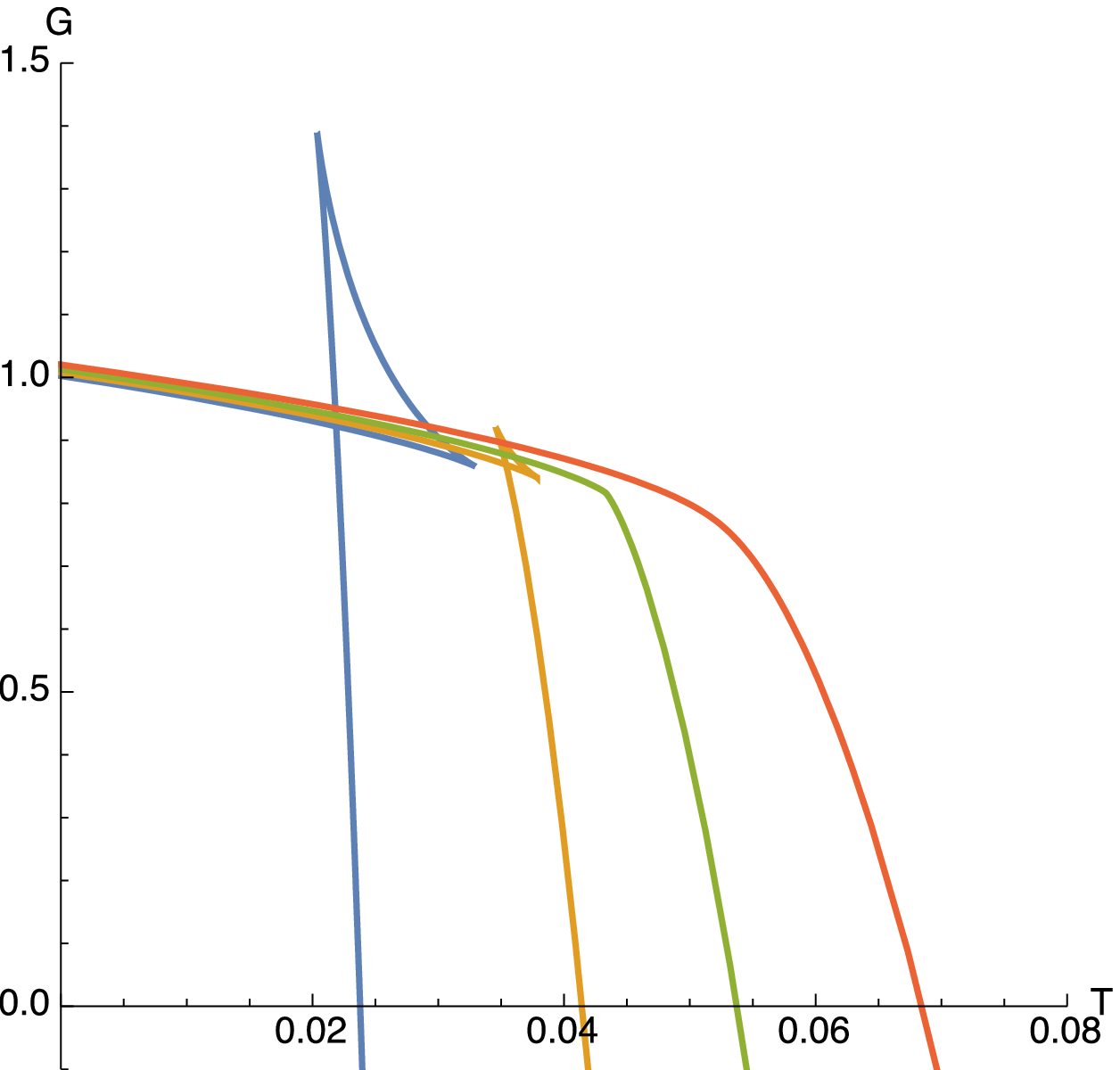}
}
\subfigure[] 
{
    \label{orbits:b}
    \includegraphics[width=0.3\textwidth]{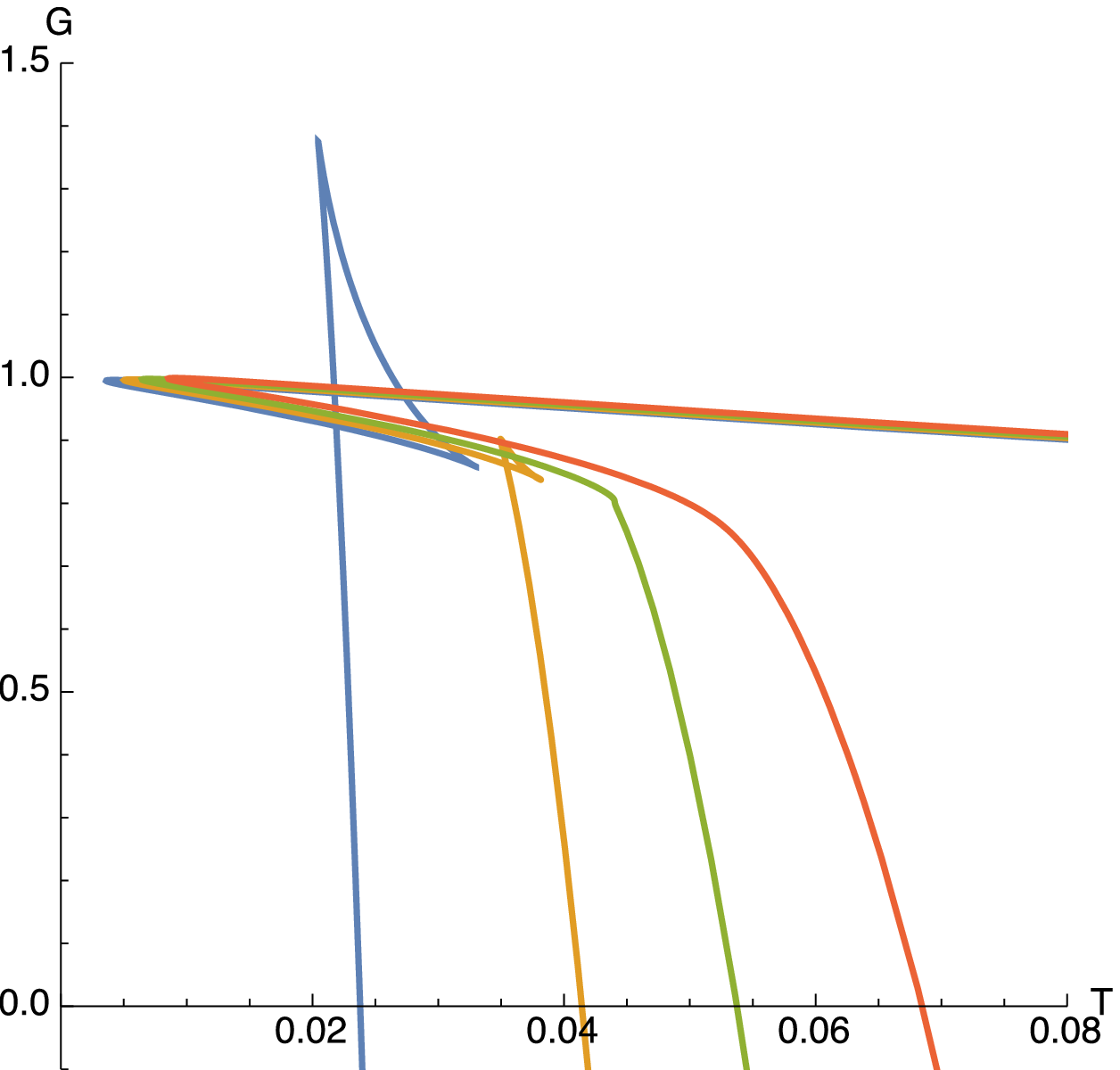}
}
\subfigure[] 
{
    \label{orbits:c}
    \includegraphics[width=0.3\textwidth]{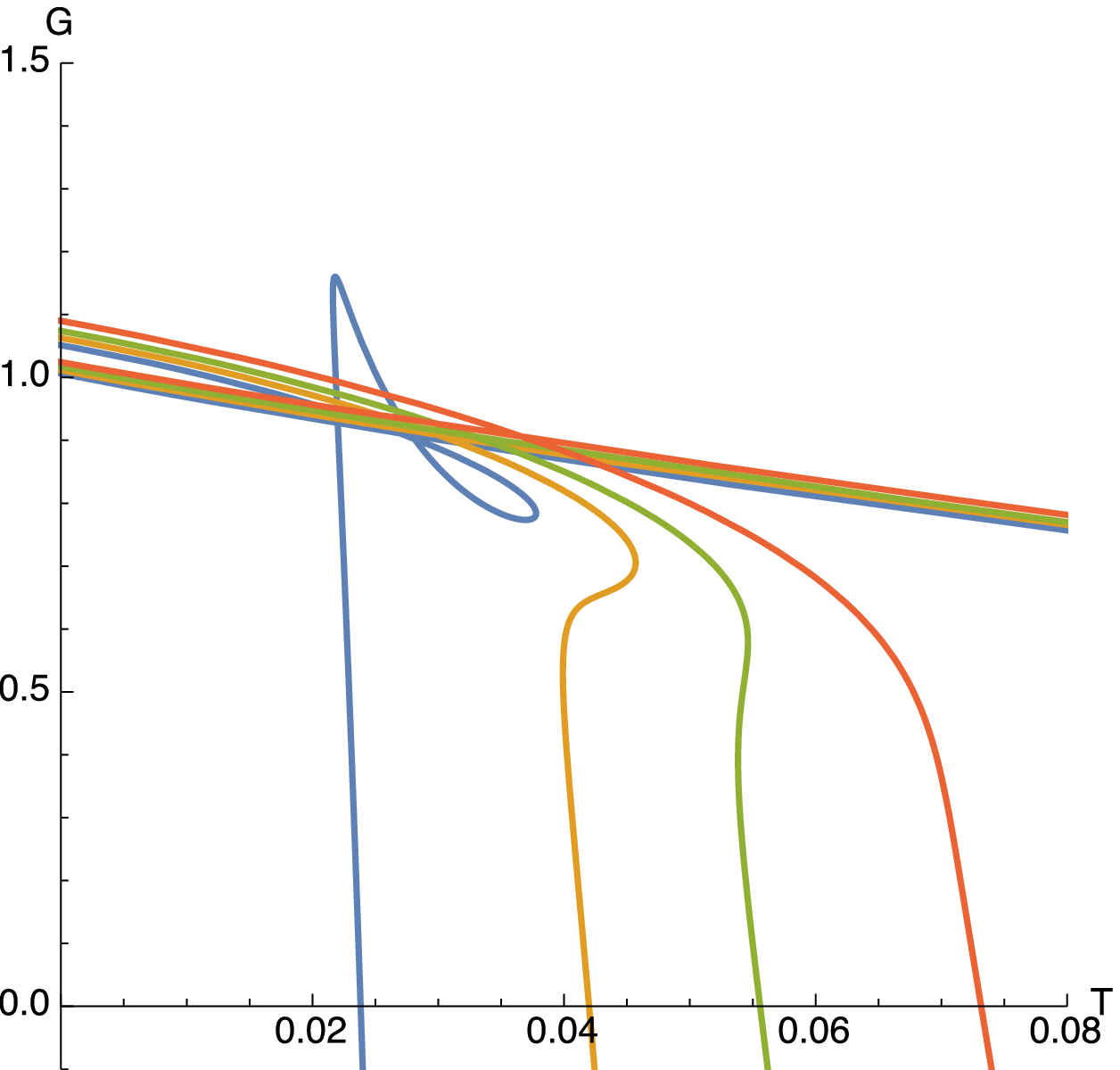}
}
\subfigure[] 
{
    \label{orbits:d}
    \includegraphics[width=0.3\textwidth]{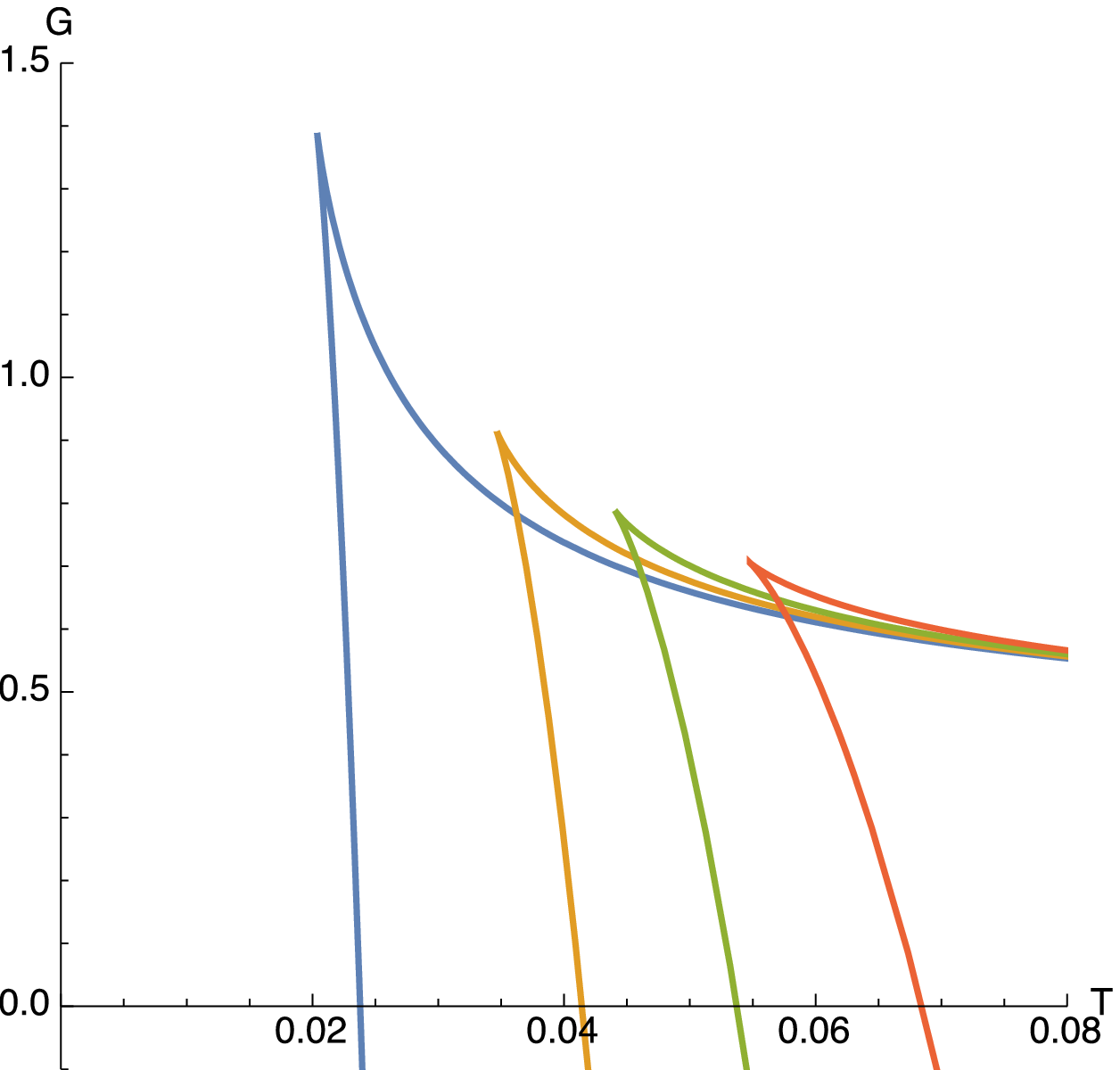}
}
\subfigure[] 
{
    \label{orbits:e}
    \includegraphics[width=0.3\textwidth]{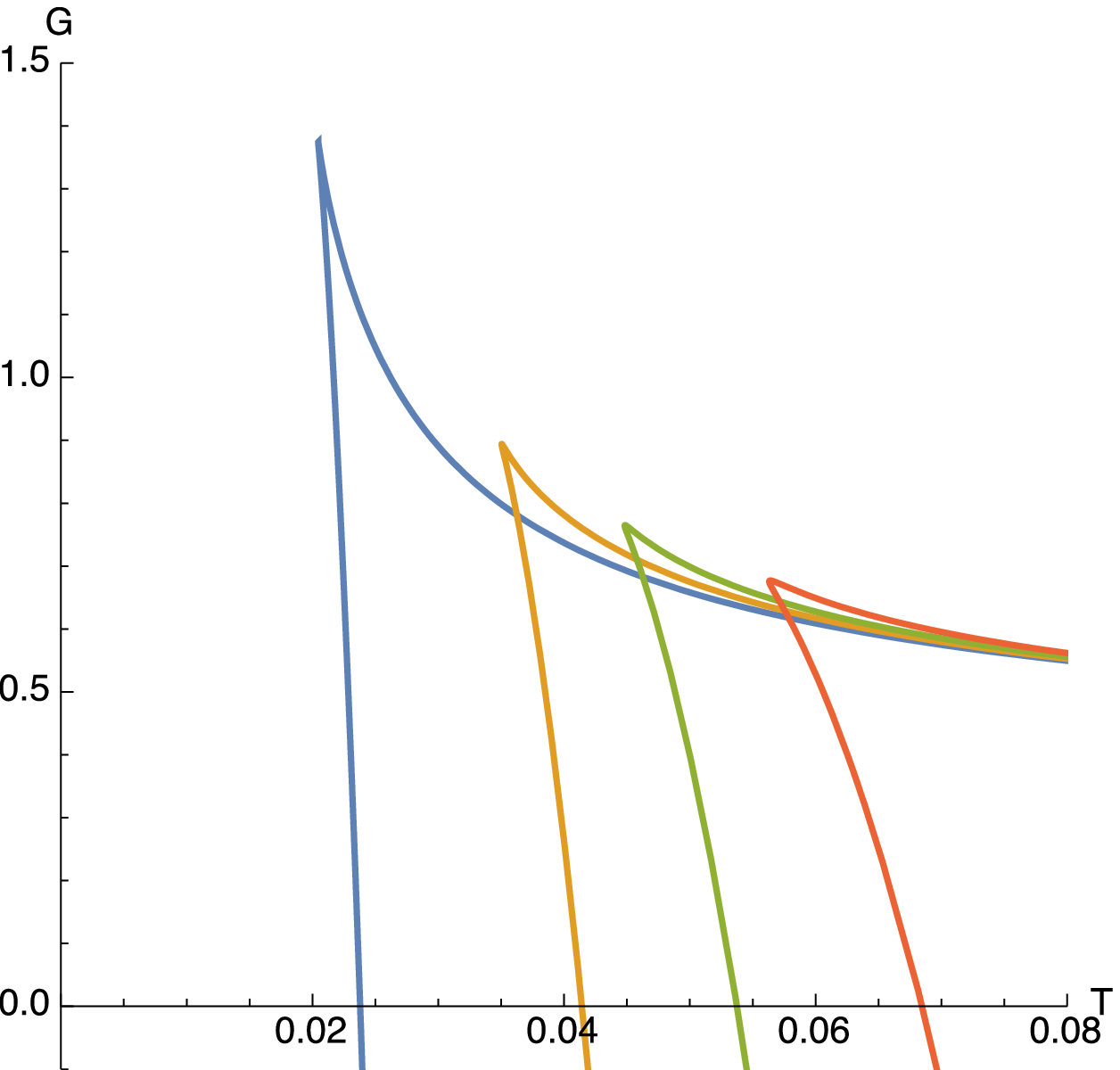}
}
\subfigure[] 
{
    \label{orbits:f}
    \includegraphics[width=0.3\textwidth]{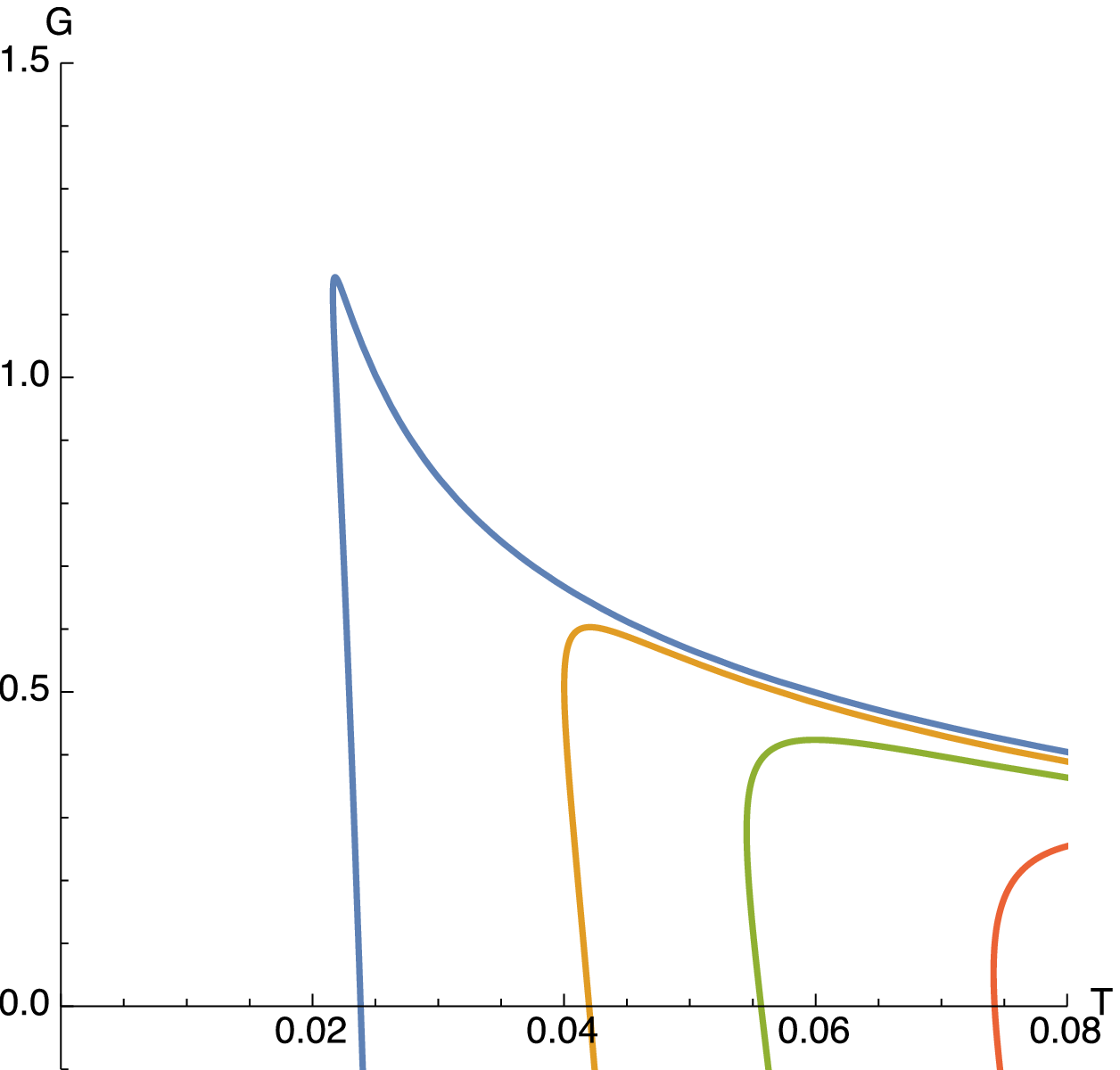}
}
\caption{The plots show the Gibbs function for different values of the noncommutative parameter $\theta$ and the BI parameter $b$: in the upper row $b = 0.001$ and $\Theta = 0, 0.1, 0.4$ from left to right, in the lower row $b = 10$ and $\Theta = 0, 0.1, 0.4$ from left to right. In each graph the pressure $p$ takes the values $96\pi p = 0.2, 0.6, 1, 1.6$ and $Q=1$. Classical RN and Schwarzschild behaviours are shown in graphs~\ref{orbits:a} and~\ref{orbits:d} respectively.}
\label{fig3} 
\end{figure}

\end{widetext}

To analyse the existence of phase transitions in the noncommutative AdSEBI black hole solution, we plot in Fig.~\ref{fig3} the noncommutative Gibbs function
\be
G = M_{ADM} - T_H S,
\ee
for several values of the noncommutative parameter $\Theta$. The explicit expression for $G$ can be obtained by using Eq.~(\ref{EC41b}) together with Eq.~(\ref{EC42}) and $S := \pi r_H^2$. The BI parameter $b$ has a nontrivial effect on the derivatives of $G$ in the commutative case, allowing the transition from Schwarzschild to RN spacetimes, and we have the standard graphs for the Gibbs function of those two situations. When $\Theta$ is not vanishing and increasing, noncommutativity manifest, and we see clearly that the curves become smoother. Eventually, the derivatives of $G$ become single-valued for $\Theta = 0.4$; a phase transition is no longer present. 

\section{Conclusions}

Using a noncommutative gauge theory of gravity based on the Seiberg-Witten map, we have constructed a noncommutative model of gravity coupled to the nonlinear BI electrodynamics. We calculated the metric coefficients associated with the AdSEBI spacetime explicitly up to second order on the noncommutative parameter using noncommutative gauge gravity fields. This approach is perturbative, the lowest order being the classical AdSEBI spacetime, and it allows the incorporation of noncommutative effects into standard results.

Furthermore, quantities like the ADM mass and the Hawking temperature of the AdSEBI black hole solution allowed quick calculations. Besides this feature, the analysis of thermodynamical properties was also straightforward. In particular, we focused on the equation of state and the Gibbs function of the noncommutative model. We showed that the most direct effect of noncommutativity on classical thermodynamical behaviour concerns the existence or absence of critical points. By looking at the Gibbs function $G$, we can determine if a phase transition may exist; it turns out that $\Theta \neq 0$ implies that the Gibbs function becomes smoother and its derivatives with respect to the temperature become single-valued, removing thus critical points; $\Theta$ does not need to take large values for this situation to happen. Our results are in agreement with previous results~\cite{Gonzalez:2015mpa} using smeared distributions of matter and charge to include noncommutative effects on classical black hole solutions.  

In~\cite{Roychowdhury:2014cva}, a formalism was put forward to explore in more detail the nature of phase transitions and to analyse features of holographic superconductors; it would be interesting to apply this technique to noncommutative inspired black holes solutions to gain more information on the influence of the noncommutative parameter on the thermodynamical properties of black holes.  

\acknowledgments
R. L. acknowledges partial support from CONACyT Grant 237351 ”Implicaciones físicas de la estructura del espacio-tiempo”. O. S-S. acknowledges support from a postdoctoral PROMEP grant.

\appendix
\section{Temperature and equation of state of the AdSEBI black hole}

For convenience, we provide here the full expression for the temperature and the equation of state of the noncommutative AdSEBI black hole. Up to second order on $\Theta$, they are

\begin{widetext}

\begin{eqnarray}
T_{H} &=& {1\over 4\pi}{d\hat g_{00}\over dr}\biggl|_{r_H}={1\over 4\pi r_H}\biggl[1+{2r_H^2\over b^2}-\Lambda r_H^2+{2r_H^2\over b^2}\biggl(1-\sqrt{1+{b^2Q^2\over r_H^4}}\biggl)\biggl]
\nonumber \\[4pt]
&&+{1\over 4\pi}\biggl[-{1\over 32}\biggl\{{4\over r_H}\biggl({2Q^2\over\sqrt{r_H^4+b^2Q^2}}-1\biggl)\biggl({1\over r_H}+{2r_H\over b^2}\biggl(1-\sqrt{1+{b^2Q^2\over r_H^4}}\biggl)-\Lambda r_H\biggl)+\biggl({1\over r_H}-\Lambda r_H
\nonumber \\[4pt]
&&+{2r_H\over b^2}\biggl(1-\sqrt{1+{b^2Q^2\over r_H^4}}\biggl)\biggl)^2\biggl\}+{1\over 16}\biggl\{{8\over r_H^3}\biggl({2Q^2\over\sqrt{r_H^4+b^2Q^2}}-1\biggl)^2+4\biggl({2Q^2\over\sqrt{r_H^4+b^2Q^2}}-1\biggl)
\nonumber \\[4pt]
&&\biggl({1\over r_H}+{2r_H\over b^2}\biggl(1-\sqrt{1+{b^2Q^2\over r_H^4}}\biggl)-\Lambda r_H\biggl)+2r_H\biggl({6\over r_H^3}+{4\over b^2r_H}-{2\Lambda\over r_H}-{4(3b^4Q^4+6b^2Q^2r_H^4+r_H^8)\over b^2r_H^3(r_H^4+b^2Q^2)^{3/2}}\biggl)
\nonumber \\[4pt]
&&\biggl({1\over r_H}+{2r_H\over b^2}\biggl(1-\sqrt{1+{b^2Q^2\over r_H^4}}\biggl)-\Lambda r_H\biggl)+2\biggl({1\over r_H}+{2r_H\over b^2}\biggl(1-\sqrt{1+{b^2Q^2\over r_H^4}}\biggl)-\Lambda r_H\biggl)\biggl({4\over r_H^2} +{4\over b^2}
\nonumber \\[4pt]
&&-2\Lambda-{4(2b^4Q^4+5b^2Q^2r_H^4+r_H^8)\over b^2r_H^2(r_H^4+b^2Q^2)^{3/2}}\biggl)\biggl\}\biggl]\Theta^2+{\cal O}(\Theta^4)
\end{eqnarray}
and
\begin{eqnarray}
P &=& {T_H\over v}-{1\over 2\pi v^2}+{b^2\over 4\pi}\biggl(1-\sqrt{1+{16b^2Q^2\over v^4}}\biggl)+{1\over 8\pi v}\biggl[{1\over 32}\biggl\{v\biggl(-{8\over v^2}+{64Q^2\over v^2\sqrt{v^4+16b^2Q^2}}\biggl)
\nonumber \\[4pt]
&&\biggl(4\pi T_H+{v\over b^2}\biggl(1-{v\over b^2}\sqrt{1+{16b^2Q^2\over v^4}}\biggl)-b^2v\biggl(1-{v\over b^2}\sqrt{1+{16b^2Q^2\over v^4}}\biggl)\biggl)
\nonumber \\[4pt]
&&+\biggl(4\pi T_H+{v\over b^2}\biggl(1-{v\over b^2}\sqrt{1+{16b^2Q^2\over v^4}}\biggl)-b^2v\biggl(1-{v\over b^2}\sqrt{1+{16b^2Q^2\over v^4}}\biggl)\biggl)^2\biggl\}
\nonumber \\[4pt]
&&-{1\over 16}\biggl\{v\biggl(-{8\over v^2}+{64Q^2\over v^2\sqrt{v^4+16b^2Q^2}}\biggl)^2+4\biggl(-{8\over v^2}+{64Q^2\over v^2\sqrt{v^4+16b^2Q^2}}\biggl)
\nonumber \\[4pt]
&&\biggl(4\pi T_H+{v\over b^2}\biggl(1-{v\over b^2}\sqrt{1+{16b^2Q^2\over v^4}}\biggl)-b^2v\biggl(1-{v\over b^2}\sqrt{1+{16b^2Q^2\over v^4}}\biggl)\biggl)
\nonumber \\[4pt]
&&+2\biggl(4\pi T_H+{v\over b^2}\biggl(1-{v\over b^2}\sqrt{1+{16b^2Q^2\over v^4}}\biggl)-b^2v\biggl(1-{v\over b^2}\sqrt{1+{16b^2Q^2\over v^4}}\biggl)\biggl)
\nonumber \\[4pt]
&&\biggl({4\over b^2}+{4\over v^2}+{16\pi T\over v}+{v\over b^2}-4b^2\biggl(1-{v\over b^2}\sqrt{1+{16b^2Q^2\over v^4}}\biggl)-{4(512b^4Q^4+80b^2Q^2v^4+v^8)\over b^2v^2(v^4+16b^2Q^2)^{3/2}}\biggl)
\nonumber \\[4pt]
&&\biggl({8\over b^2 v}+{32\over v^3}+{32\pi T\over v^2}-{8b^2\over v}\biggl(1-{v\over b^2}\sqrt{1+{16b^2Q^2\over v^4}}\biggl)-{8(768b^4Q^4+96b^2Q^2v^4+v^8)\over b^2v^3(v^4+16b^2Q^2)^{3/2}}\biggl)\biggl\}\biggl]\Theta^2+{\cal O}(\Theta^4)
\end{eqnarray}
\end{widetext}
respectively.


\end{document}